# What if Wireless Routers were Social?

*Analyzing Wireless Mesh Networks from a Social Networks Perspective*

Miray Kas, Sandeep Appala, Chao Wang, Kathleen M. Carley, L. Richard Carley, Ozan K. Tonguz

Carnegie Mellon University
Pittsburgh PA15213 USA
{mkas@ece, sappala@andrew, cw1@andrew, kathleen.carley@cs, carley@ece, tonguz@ece}.cmu.edu

*Abstract*—Wireless Mesh Networks (WMNs) consist of radio nodes organized in a mesh topology for serving wireless mesh clients to communicate with one another or to connect to the Internet. Nodes in a mesh network can communicate with each other either directly or through one or more intermediate nodes, similar to social networks. WMNs share many common properties with social networks. We first identify the differences and similarities between social networks and WMNs and then use metrics that are typically used for social network analysis (SNA) to assess real WMNs. Analyzing real WMN data collected from the UCSB MeshNet and MIT Roofnet testbeds reveals that using SNA metrics are helpful in designing WMNs with better performance. We demonstrate the validity of our conclusions and this new approach by focusing on two sample applications of social networks: network reliability assessment and channel access scheduling.

## I. INTRODUCTION

A wireless mesh network (WMN) is a multi-hop communication network in which the nodes (routers) are self-organized (*i.e.*, without needing a central coordinator) to form a mesh topology to provide communication over multiple wireless links without necessarily requiring an external authority imposing a planned structure. Over the last decade, wireless mesh networking technology has emerged as an important enabling technology to provide better services in wireless networks. As in all types of wireless networks, bandwidth is a very scarce resource. Improving the performance of multi-hop wireless mesh networks is currently a very active research area.

A social network (SN) is a social structure consisting of a group of people that are connected by various relationships such as friendship, family ties or common interests and beliefs. Social networks are traditionally modeled and analyzed as graphs where the social actors (*i.e.,* people) are represented as nodes while the relationships between the people are represented by the links drawn between the nodes on these social network graphs.

Research on social networks in the past 60 years or so has led to a wealth of findings about the structure and evolution of these networks and a host of metrics and tools for assessing, forecasting, and visualizing network behavior more generally. Historically, much of the relevant work used social networks analysis (SNA) in which the behavioral patterns and social interactions among human beings were assessed using graph theoretic metrics. Traditional SNA is mostly performed on static snapshots of the network, targeting small networks such as networks observed among a class of students or within a study group.

Between WMNs and social networks, there exist both similarities and differences which enable one to draw certain analogies while making it hard to get a one-to-one mapping between these two types of networks. For instance, the functions and existence reasons of WMNs and social networks among humans are very different. Like all man-made communication networks, WMNs are not affected by emotions as human networks are.

On the flip side of the coin, however, there exist many similarities that encourage us to draw analogies that are primarily borne from applicability of graph-theoretic representations for both types of networks. For instance, while the links in both social networks and WMNs may exhibit bursts of activity, the overall pattern of the connections among nodes, on average, is fairly static. In social networks, the significance or the frequency of the relations among agents are usually represented as link weights. Wireless links also have link-weight information such as SNR, showing the links' quality of communication. Similarly, both WMNs and social networks can be multi-modal, modeling links that represent different types of relationships. For instance, in social networks, the same set of people can be modeled by both family and friendship ties. In WMNs, this corresponds to multi-channel communication where each node is equipped with multiple Network Interface Cards (NICs) operating on different channels [1].

One important property that both social networks and wireless networks have in common is the graph based description of both types of networks, which in turn allows for applying similar mathematical tools and approaches to both network types. For instance, a feature that is frequently observed in both types of networks is transitivity although it is not the only factor contributing to the formation of the final topology of networks. In social networks, transitivity refers to the probability of agents *i* and *k* being friends given that there exist friendship ties (*i*, *j*) and (*j*, *k*). Since the links in WMNs are affected by the coverage areas of routers, it is very likely that nodes will have common neighbors, forming many triangles of the triplet (*i*, *j*, *k*).

In this paper, inspired by the synergy that can be observed between these two types of networks, we propose using social network metrics to identify the nodes that are crucial in a

WMN by exploiting the analogies that can be drawn between WMNs and traditional static social networks. We also demonstrate how social network metrics can be utilized in WMNs via two case studies: *(i)* reliability assessment and *(ii)* channel access scheduling.

## II. BACKGROUND ON SOCIAL METRICS

SNA centrality measures focus on finding the key actors in a social network. There are various metrics proposed for evaluating the prominence/importance of the actors in the network from different aspects. The most well-known ones are either degree based or geodesic distance based metrics.

Table 1 lists the most commonly used key actor metrics in social network analysis. Depending on the research question at hand, one centrality metric might become more important than the others because each metric provides insights into different aspects of the networks and each has different implications and usages.

**Table 1-Common centrality metrics in social network analysis.**

| Measure | Definition | Usage |
| --- | --- | --- |
| Degree Centrality | Node with most connections | Identifying sources for intel |
| Betweenness Centrality | Connection between disconnected groups | Reducing activity by disconnecting groups |
| Closeness Centrality | Node that is closest to all other nodes | Rapid access to information |
| Eigenvector Centrality | Nodes connected to most highly connected nodes | Identifying who can mobilize others |

Degree based metrics consider the number of connections a node has. *Degree centrality* of a node is simply defined as the number of its connections. Another degree based centrality measure, *spectral centrality*, is a recursively calculated metric which defines an actor/node as prominent if it is pointed to by another prominent actor. *Eigenvector centrality* is another metric used for key actor identification, which defines the centrality value of a node to be proportional to the sum of centrality values of all its neighbors. In other words, it is used for finding the node that is most connected to other highly connected nodes, indicating a stronger capital.

Unlike degree based metrics, geodesic distance based metrics focus on the network topology, the connections, and the distances between the nodes. *Closeness centrality* of a node evaluates its information propagation efficiency. For any *node-X*, closeness centrality is defined as the inverse of the sum of distances between *node-X* and all other actors in the network. *Betweenness centrality* is defined as the number/fraction of the best (shortest) paths that pass through a *node-X*. For instance, in a clustered network, a node that is high in betweenness is likely to be a node that connects two clusters. Another centrality metric derived from betweenness centrality is *bridging centrality*. Bridging centrality of a *node-X* is calculated by multiplying its betweenness centrality value by a bridging coefficient such that it indicates how well the *node-X* is positioned among nodes with high degree centralities.

We next formulate the centrality metrics used in this paper.

**Total Degree Centrality** of node $i$ is loosely defined as the number of its immediate neighbors. The nodes that have higher degree centrality have more connections to others in the network.

$$Total\_Degree(i) = \frac{\sum_{j \in N} x_{ij}}{n-1} \text{ where } x_{ij} = 0 \text{ or } x_{ij} = 1$$

This metric corresponds to the traditional 1-hop neighborhood size that is commonly used in wireless networks, scaled by the number of nodes *n* in the network.

**Closeness Centrality** of node $i$ describes its efficiency of information propagation to all others. It is defined as the inverse of the average of the distances between $i$ and all other nodes in the network. When two nodes $i$ and $j$ are not connected $distance(i,j) = \infty$, and $distance(i,j)$ is not included in the computation of closeness centrality for node $i$.

$$Closeness(i) = \frac{n-1}{\sum_{j \neq i} distance(i,j)}$$

**Betweenness Centrality** of node $i$ is defined as the percentage of shortest paths across all possible pairs of nodes that pass through node $i$. Let $g_{j,k}$ be the number of shortest paths in from $j$ to $k$ and $g_{j,k}(i)$ be the number of shortest paths from $j$ to $k$ that contain $i$.

$$C_B = \sum_{j<k} \frac{g_{j,k}(i)}{g_{j,k}} \text{ where } i \neq j, i \neq k$$

The value of $C_B$ is then normalized by the number of possible node pairs to calculate the betweenness centrality of $i$.

## III. RELATED WORK ON WMN-SNA INTERFACE

Historically, social networks have been well studied by disciplines such as anthropology, sociology, psychology, and business management. Only recently have wireless network researchers realized the significant amount of work that has been done for social network analysis and started borrowing techniques/metrics from social network analysis (SNA) to design better networking protocols for wireless ad-hoc, mesh, and delay-tolerant networks [2].

To exemplify a few, [3] adjusts nodes' willingness to forward data on behalf of other nodes according to their approximate bridging centrality values. In another paper, [4], the authors extend their proposal in [3] and introduce a load-aware version of this metric.

Betweenness centrality is another metric that is used in a number of wireless network papers. For instance, [5] uses betweenness centrality to perform caching in wireless sensor networks while [6] uses it for multicasting in delay tolerant networks. 'Delay tolerant networks' is a subfield of wireless networks that has explored social network concepts the most [6], [7], [8].

## IV. WIRELESS MESH NETWORK (WMN) DATASETS

In this paper, we use the datasets provided by WMN deployments: UCSB MeshNet' and 'MIT's Roofnet'. To perform social analysis on these networks, we use ORA [9] which is an interactive network analysis tool that maintains the internal structure of an organization/social network as a set of agents, tasks, and resources.

The UCSB MeshNet is a multi-radio 802.11 a/b network consisting of 38 PC-nodes deployed indoors on five floors of a typical office building in the UCSB campus. The data contains 2 sub-networks, each consisting of 19 nodes.

The MIT Roofnet consists of 22 nodes spread over four square kilometers in Cambridge, MA. Each node is a PC equipped with a Prism2-chipset 802.11b radio and an omnidirectional antenna that is either roof-mounted or projecting out of a window. All radios operate on the same 802.11b channel.

## V. NODE-LEVEL ANALYSIS

In an attempt to understand the relative importance of the WMN nodes in the network and identify potential roles nodes in a WMN can be assigned, we perform node-level analysis on a subnet of the UCSB Meshnet (Figure 1).

Table 2 shows the five top-ranked nodes in terms of degree, closeness, and betweenness centrality along with their corresponding values presented in parentheses.

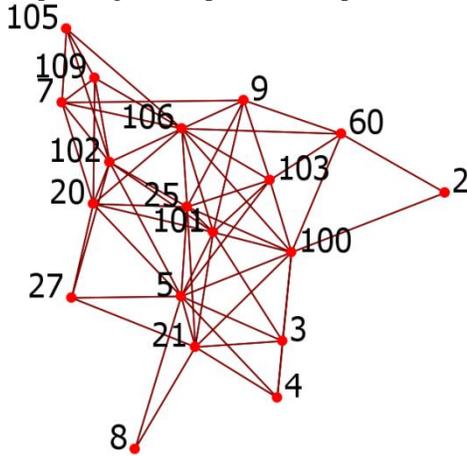

**Figure 1-UCSB Meshnet Topology (19-node subnetwork). All node labels in the figure are preceded with 10.2.1. to form their IP addresses.**

**Table 2 -Social Centrality Rankings of Nodes.**

| Degree Centrality | Closeness Centrality | Betweenness Centrality |
|---|---|---|
| 10.2.1.5      (0.556) | 10.2.1.5      (0.720) | 10.2.1.5      (0.176) |
| 10.2.1.106 (0.556) | 10.2.1.106 (0.692) | 10.2.1.100 (0.153) |
| 10.2.1.102 (0.500) | 10.2.1.102 (0.643) | 10.2.1.106 (0.146) |
| 10.2.1.102 (0.500) | 10.2.1.101 (0.643) | 10.2.1.102 (0.107) |
| 10.2.1.101 (0.500) | 10.2.1.20   (0.621) | 10.2.1.101 (0.086) |

Since the topology shown in Figure 1 is a well-connected network, similar nodes are selected as the most central nodes by different centrality metrics. Yet, different metrics rank/distinguish them differently. For instance, nodes 10.2.1.5 and 10.2.1.106 are the top two nodes for closeness and degree centrality. Both nodes have the same degree (*i.e.*, their 1-hop neighbor counts are the same), however, 10.2.1.5 is able to access other resources more efficiently than 10.2.1.106, given their closeness values.

## VI. IMPLICATIONS OF SOCIAL ANALYSIS ON WMNs

In this section, we discuss a high-level set of example use cases for utilizing social network analysis in designing higher performance wireless networks.

**Failure Detection.** Providing seamless operation to end users requires fast detection and recovery of link and/or node failures. Nodes with high closeness centrality values can be useful for recovery updates in the network. Since closeness centrality measures how close a node is to all other nodes in a network, nodes with high closeness values can sense and access most of the nodes in the network very rapidly, incurring shorter delays.

**Routing.** Shortest path related metrics are widely studied for routing and monitoring purposes in wireless networks. Recently, a routing algorithm –which is primarily designed for delay tolerant networks– that is based on exchanging pre-estimated betweenness centrality values and locally determined social similarity (based on the number of common neighbors) to the destination node has been proposed [8]. When the destination node is unknown to the current sender, routing request messages can be directed towards more central nodes in the hopes of finding the destination sooner.

**Multicast Operations.** Multicasting is a promising method that can be adopted in WMNs for reducing bandwidth consumption of many applications and services running on the network [10]. Nodes with higher Eigenvector centrality values can be useful for multicasting purposes because such a node is central to the extent that its neighbors are also central. Such a design would reduce flooding in the network since the required message is received by the rest of the network in fewer steps.

**Multi-Radio MAC.** Multi-radio Unification Protocol (MUP) is one of the commonly used MAC layer protocols in networks with multi-radio nodes because it coordinates the operation of wireless network cards on non-overlapping channels and it allows every Tx/Rx to communicate independently in a different channel [11]. Wireless network interface cards send single-hop probe messages to check link quality before transmission. Such probe messages can include centrality values of nodes to direct traffic towards central nodes, hence to get higher performance. Although multi-radio MAC protocols is one potential application of social network analysis on wireless networks, this scenario is less likely than the other scenarios discussed in this section.

**Network Management.** Wireless network management is a broad term that covers numerous research and engineering problems exploring concerns such as configuration, provision, diagnosis, or optimization of wireless (mesh) networks. Social centrality metrics can aid system administrators or automated management systems to better analyze the state of a WMN, and manage it in a more effective manner [3]. Social centrality metrics provide answers for questions like: *(i)* Which nodes are more critical from a robustness point of view? *(ii)* Loss of which nodes would have a significant impact on the connectivity of the network? In Section VII, we analyze this case further, and discuss various node failure scenarios.

**Network Partitioning.** Network partitioning refers to cutting networks into internal subcomponents (domains). Then filtering policies can be applied on these separate domains to achieve security enhancement and flow control. Nodes that have high betweenness centrality are likely to be one of those potential cut-points as they are on a high fraction of shortest paths among other nodes in the network. Hence, nodes with higher betweenness centrality values can be used to partition the network into sub-networks and serve as gateways between different components.

**Channel Access Scheduling.** The MAC protocols that are available in the literature can be broadly classified into two groups: contention based protocols and scheduling based protocols. In contention based protocols, nodes contend for channel access and collisions are possible. 802.11 MAC protocol [12] which is based on carrier sense multiple access/collision avoidance (CSMA/CA) is one of the most well-known examples of contention based MAC protocols. The second group of MAC protocols, the scheduling-based protocols, schedules the access of nodes or links to the channel in advance. TDMA based protocols that operate in discrete slotted time and typically arrange the transmission of the nodes or links in the network based on a schedule constitute examples of scheduling-based protocols.

The relative importance of WMN nodes can be incorporated into channel access scheduling algorithms. Node priorities can be adjusted according to their centrality values such that central nodes are assigned a higher number of time slots (TDMA) or use smaller windows for exponential random backoff (CSMA). In Section VIII, we propose a cross-layer, distributed channel access scheduling scheme that exploits closeness centrality of nodes for prioritization.

## VII. CASE STUDY - I: COORDINATED ATTACKS

In this section, we investigate the impact of social network analysis on reliability assessment. We perform coordinated attacks (*i.e.,* introduce failures to the central nodes) and discuss the impact of social centrality metrics in terms of the average number of hops packets travel in the network.

**Failure Scenarios:** For each centrality metric of interest (e.g. betweenness centrality, closeness centrality, and degree centrality), we progressively select up to the first five nodes with highest centrality from the UCSB Meshnet.

We configure the selected nodes to incorporate a statistical error model (e.g., uniform random error) over incoming wireless channels. The incoming error module lets each receiver experience packet corruption with different degrees of error since the error is independently computed for each error module. To model node failures without actually removing the nodes from the ns-2 topology and changing the total amount of traffic generated, we set the error rate to 1 for the selected high-centrality nodes.

**Traffic Pattern:** For each experiment, we simulate the same uniform traffic scenario where every node generates a CBR connection to every other node, resulting in $O(n^2)$ connections. All the connections start at the 25th second and end at the 125th second and the CBR rate is fixed at 500 bps for all connections. The simulations last for 200 seconds.

**Routing Protocol:** In our simulations, we use Optimized Link State Routing protocol (OLSR) as our routing protocol. It is a proactive link state routing protocol where at each node next-hop destinations for all nodes in the network are maintained using shortest-hop paths. In the case of node removals, its behavior is consistent across different runs on the same network topology (e.g., when node *x* is removed in two different runs, newly formed shortest paths are consistent across different runs.)

Figure 2 shows the impact of central nodes' failures on the average number of hops packets traverse to reach their destinations. *Random*, our baseline, shows the average of 10 experiments where the failing nodes are selected randomly.

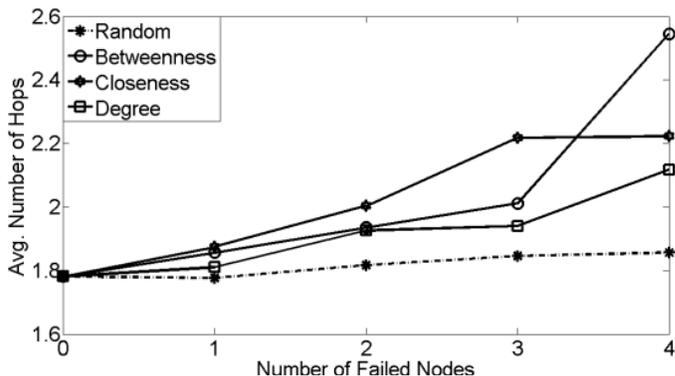

**Figure 2 - Increase in average number of hops as top ranked nodes are progressively removed.**

While closeness centrality identifies nodes that have rapid access to information by being close to many other nodes on average, betweenness centrality detects nodes that are on the shortest path for many other nodes which are usually the nodes that can partition the network. Therefore, in Figure 3, when 1 or 2 nodes fail, the impact of betweenness centrality is less than that of closeness centrality because the original topology is relatively well connected and it is not immediately partitionable. However, as the number of failing nodes increases, the residual topologies have longer paths causing the steep increase in the betweenness centrality results.

Another interesting point is that degree centrality, the social centrality metric corresponding to the 1-hop neighborhood size frequently used in wireless research, is not as effective at identifying critical nodes as other social centrality metrics. The increase in the average number of hops caused by degree centrality is consistently lower than those of closeness and betweenness centrality. This is because degree centrality is not related to shortest paths in the network while closeness and betweenness are. Delivery of packets in wireless mesh networks heavily rely on the shortest paths identified by the routing algorithms, which are easily targeted by geodesic distance based centrality metrics such as betweenness and closeness, while not necessarily by degree based centrality metrics.

To show that our results are generalizable to large-scale networks, we perform simulations on a 200-node network as well, and progressively remove up to 40 nodes (20%). The results in Figure 3 justify the relative ranking of centrality metrics to be betweenness, closeness and degree centrality in terms of their importance for network reliability. In other words, the metric that is most effective in degrading network performance in a coordinated attack is betweenness centrality. This behavior is consistent with other work on the literature that assesses the effectiveness of centrality metrics in coordinated attacks [13] .

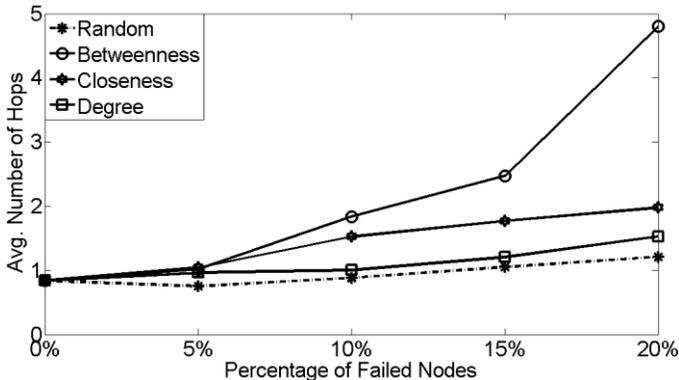

Figure 3 - Increase in average number of hops as top ranked nodes are progressively removed from a 200-node topology.

VIII. CASE STUDY - II: SOCIALLY - AWARE TDMA

Recent WMN standards such as WiMAX [14] and 802.11s [15] consider Spatial-TDMA (STDMA) based MAC mechanisms and WMNs operate in multi-hop environments. Hence, in this paper, we focus on STDMA based schemes at the MAC layer. In STDMA based schemes, two nodes that are in non-conflicting parts of the network can be scheduled to transmit simultaneously.

We propose a STDMA-based distributed cross-layer channel access scheduling scheme based on social network analysis. We use *closeness* as our social centrality metric and prioritize wireless medium accesses of nodes that are ranked higher in terms of closeness over other nodes.

We prefer using closeness centrality because, by definition, it is used to describe information propagation efficiency and it is an appropriate metric for optimizing the efficiency of communication networks, including WMNs. In addition, the computation of closeness requires fewer resources compared to other social centrality metrics because Eigenvector centrality has a recursive implementation and betweenness requires information on all shortest paths in the network.

Another concern about the use of betweenness values for a prioritization scheme stems from the distribution of betweenness values in larger networks. Betweenness value of a node indicates the fraction of shortest paths it is on across all possible node pairs in a network. However, not all nodes can be on many shortest paths. In the literature, betweenness values of nodes in large networks have been shown to follow power-law distributions where a small number of nodes have high betweenness values while many others are zero [16]. In this case, a prioritization scheme based on betweenness values is unable to distinguish many nodes that have zero betweenness. In addition, in the case of scheduling, it may lead to starvation of nodes with 0.0 betweenness centrality unless there is a starvation prevention mechanism in place.

### A. Cross-Layer Dissemination of Centrality Values

For the routing layer, we use Optimized Link State Routing protocol (OLSR). Since it is a link state protocol, each node maintains network topology information in their routing tables, enabling calculation of the social centrality metrics.

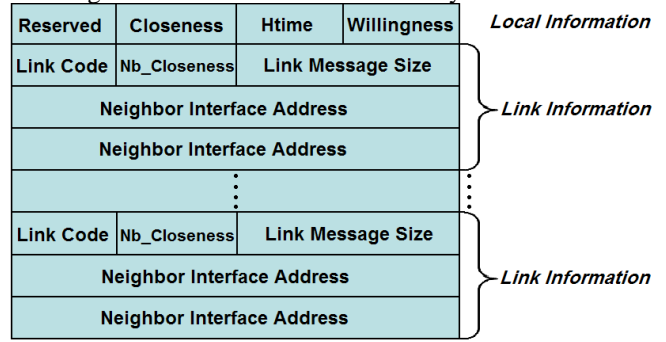

Figure 4 - Extended OLSR Hello message.

For the dissemination of closeness values, we utilize periodical HELLO messages broadcast by each node every 2ms. A HELLO message contains information about the originating node and advertises its links to its 1-hop neighbors. We extend HELLO messages to include the originating node's and its 1-hop neighbors' closeness values so that all nodes can learn the closeness values of all nodes within their 2-hop neighborhood. In particular, we replace the unused reserved fields with 'Closeness' field for the originating node, and 'Nb_Closeness' field for the advertised 1-hop nodes as shown in Figure 4. Therefore, no additional control messages are required for exchanging closeness priorities.

### B. Socially-Aware TDMA Scheduling

We propose Socially-Aware TDMA channel access scheduling algorithm that performs a lottery based slot assignment where the nodes' closeness values are used as their approximate priorities. We aim to improve throughput by assigning more slots to more central nodes.

---

**Algorithm-1: Socially-Aware TDMA Scheduling**

**Data:** Topology and closeness information for 2-hop neighborhood of node *i*, and current *FrameCount*.
**Result:** The set of time slots node *i* is eligible to transmit during the next frame.
for $j \leftarrow 1$ to FRAME_SIZE do
    *slotID* ← FormSlotID (*FrameCount, j*);
    *localLst* ← FormLocalTicketsByHash (*i, slotID*);
    *nbrLst* ← FormNeighborTicketsByHash (*i, slotID*);
    *contenders* ← *nbrLst* ∪ *localLst*;
    *winner* ← FindMaxInMeshElection (*slotID, contenders*);
    if *localLst*.Contains(*winner*)
        *i*.slots[*j*].status ← WON;

---

We divide the execution time into slots where each frame contains FRAME_SIZE many slots. At the end of each frame,

each node independently runs the distributed scheduling algorithm shown in Algorithm-1. Each node generates as many pseudorandom lottery ticket numbers as its closeness value for each time slot in the frame. Lottery tickets are pseudo-randomly generated by a simple hash function that contains only arithmetic operations and takes NodeID and SlotID as input.

This way, each node can generate unique and predictable ticket numbers for the given time slot/frame. Since all nodes run the same algorithm, each node is able to guess what its neighbors will generate as lottery ticket numbers because each node is fully aware of its 2-hop neighbors' inputs. To give an example, if a node's closeness value is 10, then it joins the elections with 10 tickets. If another node has closeness value 2, it joins the elections with 2 tickets. The node that has the highest ticket number for that slot is the winner of the slot and is the node that has the right to transmit and a node with a higher number of tickets has a higher chance of winning. With this kind of scaling, the probability of each node to win a slot will be approximately proportional to its closeness priority.

WMNs have fairly static topologies. Therefore, in our simulations, we simulate only static topologies; however, the proposed socially-aware MAC scheme is able to handle mobility. Because we use periodical OLSR HELLO packets which are broadcast every 2ms, node mobility and topology changes accounted for in the closeness calculations in real time.

### C. Performance Evaluation

Next, we evaluate the performance of the proposed socially-aware scheduling scheme. For our baseline scheme, we use the same framework with only the prioritization scheme modified. Basically, our baseline case is again a multi-hop, STDMA-based MAC scheme. In the baseline case, each node generates a random number of lottery tickets, rather than closeness many tickets. In other words, each node has an equal chance of being the winning node that earns transmission right during each time slot. And, at each time slot the winning node is selected randomly (e.g. nodes generate random weights instead of using their closeness values as their weights), following no particular prioritization.

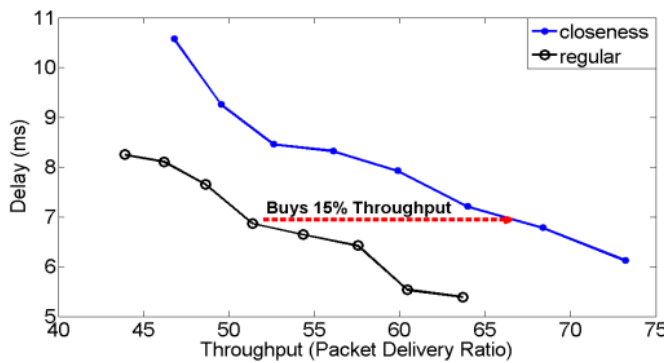

**Figure 5 - Delay vs. Throughput.**

We perform simulations in ns-2.31 using data rates from 650 bits/sec up to 1350 bits/sec. We measure end-to-end delay and end-to-end throughput calculated across all data packets (excluding control packets) generated during the simulations.

Figure 5 presents our initial results on the tradeoff between delay and throughput, justifying the throughput advantage brought by the use of social centrality metrics.

### IX. CONCLUSION

In this paper, we propose to use social network analysis for modeling wireless mesh networks. Our case study on performing coordinated network attacks shows that the failure of socially central nodes result in a substantial increase in the average number hops; highlighting the importance of using social parameters in WMNs instead of traditional 1-hop neighborhood size metric. Our simulation results for the proposed socially-aware channel access scheduling has also proved beneficial, showing that utilizing socially central nodes provides substantial throughput improvement. Future work should include extensions and performance comparisons on other well-known channel access techniques such as IEEE 802.11 and Slotted Aloha, and explore alternative size and structured topologies, and disruption metrics.

### X. ACKNOWLEDGEMENTS

This work is supported in part by the Office of Naval Research (ONR MURI N000140811186) and by the center for Computational Analysis of Social and Organizational Systems (CASOS). The views and conclusions contained in this document are those of the authors and should not be interpreted as representing the official policies, either expressed or implied, of the Office of Naval Research or the U.S. government.

**Miray Kas** (mkas@ece.cmu.edu ) has obtained her B.Sc. and M.Sc degrees in computer engineering from Bilkent University, Ankara, Turkey. Since 2009, she has been pursuing the PhD degree in the Electrical and Computer Engineering Department of Carnegie Mellon University, Pittsburgh, PA. Her current research interests are in the areas of algorithm design and trend analysis for dynamic social networks and wireless networks. Her previous publications focus on channel access scheduling for wireless mesh networks and on-chip networks.

**Sandeep Appala** (sappala@andrew.cmu.edu) received his Bachelors degree in telecommunications from R. V. College of Engineering, Bangalore India and M.Sc degree in information technology and mobility from Carnegie Mellon University. He has previously worked as a R&D Engineer on 3G/4G technologies in the Radio access division at Nokia Siemens Networks, India. In the fall of 2011, he worked at Sprint Applied research labs, Burlingame, California as a research intern on Mobile analytics and mining large scale graphs using Hadoop.

**Chao Wang** (cw1@andrew.cmu.edu) received a BS degree in Electrical Engineering from Polytechnic Institute of New York University and a BS degree in Communication Engineering from Southwest Jiaotong University, China. He is currently pursuing his Master of Science degree in Electrical and Computer Engineering at Carnegie Mellon University. His main interests are computer system, storage system.

**Kathleen M. Carley** (kathleen.carley@cs.cmu.edu) is a tenured full professor in the Institute for Software Research Department in the School of Computer Science of Carnegie Mellon University (CMU) , Pittsburgh, PA. She currently leads substantial research efforts in the areas of dynamic network analysis and information diffusion, develops new algorithms and technologies for this area ranging from text mining, to network and visual analytics, to agent-based models. She has published widely with over 350 articles in the areas of network science, organizations, simulation, and social change. Her tools, in particular AutoMap and ORA are used in a variety of settings to extract and analyze networks (see www.casos.cs.cmu.edu/tools).  In 2008, she founded with L. Richard Carley, the CMU startup known as Netanomics (www.netanomics.com) which specializes in providing technologies to support complex socio-technical systems from a combined social and technical perspective.  Her current research interests include information dynamics, social media, dynamic networks, extracting networks from massive data, complex systems, re-usable and interoperable simulations, organizational design, WMD deterrence, remote detection of CBRNE capability, and security. She currently serves, or has served, as a consultant for several companies, government agencies, and on multiple national research council panels.

**L. Richard Carley** (carley@ece.cmu.edu) received an S.B. in 1976, an M.S. in 1978, and a Ph.D. in 1984, all from the Massachusetts Institute of Technology.  He joined Carnegie Mellon University in Pittsburgh, PA in 1984, and in March 2001, he became the STMicroelectronics Professor of Engineering at CMU.  His research interests include analog and RF integrated circuit design in deeply scaled CMOS technologies and novel nano-electro-mechanical device design and fabrication. He has been granted 15 patents, authored or co-authored over 120 technical papers, and authored or co-authored over 20 books and/or book chapters. He has won numerous awards including Best Technical Paper Awards at both the 1987 and the 2002 Design Automation Conference (DAC).  In 1997, Dr. Carley co-founded the analog electronic design automation startup, Neolinear, which became part of Cadence in 2004. In 2001, Dr. Carley co-founded a MEMS sensor IC startup which morphed into a MEMS RF IC startup in 2005, and in 2007 he co-founded a Network Sciences Company - Netanomics.

**Ozan K. Tonguz** (tonguz@ece.cmu.edu) is a tenured full professor in the Electrical and Computer Engineering Department of Carnegie Mellon University (CMU), Pittsburgh, PA. He currently leads substantial research efforts at CMU in the broad areas of telecommunications and networking. He has published about 300 papers in IEEE journals and conference proceedings in the areas of wireless networking, optical communications, and computer networks. He is the author (with G. Ferrari) of the book Ad Hoc Wireless Networks: A Communication-Theoretic Perspective (Wiley, 2006). In December 2010, he founded the CMU startup known as Virtual Traffic Lights, LLC, which specializes in providing solutions to acute transportation problems using vehicle-to-vehicle and vehicle-to-infrastructure communications paradigms (http://www.ece.cmu.edu/~tonguz/vtl/). His current research interests include vehicular ad hoc networks, wireless ad hoc and sensor networks, self-organizing networks, smart grid, bioinformatics, and security. He currently serves or has served as a consultant or expert for several companies, major law firms, and government agencies in the United States, Europe, and Asia.